\documentclass[aip,rsi,amsmath,amssymb,reprint,noeprint]{revtex4-1}

\usepackage{graphicx}
\usepackage{dcolumn}
\usepackage{multirow}
\usepackage{ragged2e} 
\usepackage{gensymb} 
\usepackage{amsmath,amssymb} 
\usepackage[inline]{enumitem} 

\newcommand{\Vang}{$V_{\mathrm{ang}}$}
\newcommand{\Vtor}{$V_{\mathrm{tor}}$}
\newcommand{\rbij}{$\mathbf{r}_{ij}$}
\newcommand{\rij}{$r_{ij}$}
\newcommand{\rcut}{$r_{\mathrm{cut}}$}
\newcommand{\vepab}{$\varepsilon_{\alpha\beta}$}
\newcommand{\slj}{$\sigma_{\mathrm{LJ}}$}
\newcommand{\elj}{$\epsilon_{\mathrm{LJ}}$}
\newcommand{\sang}{$\sigma_{\mathrm{ang}}$}
\newcommand{\stor}{$\sigma_{\mathrm{tor}}$}
\newcommand{\phiab}{$\phi_{\alpha\beta}$}

\newcommand{\phiaboff}{$\phi^{\mathrm{offset}}_{\alpha\beta}$}

\begin{document}
	
	
	\title{Programming patchy particles to form complex periodic structures}
	\author{Daniel F. Tracey}
	\affiliation{Physical and Theoretical Chemistry Laboratory, Department of Chemistry, University of Oxford, South Parks Road, Oxford, OX1 3QZ, United Kingdom}
	\author{Eva G. Noya}
	\affiliation{Instituto de Qu\'{i}mica F\'{i}sica Rocasolano, Consejo Superior de Investigaciones Cient\'{i}ficas, CSIC, Calle Serrano 119, 28006 Madrid, Spain}
	\author{Jonathan P. K. Doye}%
	\email{jonathan.doye@chem.ox.ac.uk}
	\affiliation{Physical and Theoretical Chemistry Laboratory, Department of Chemistry, University of Oxford, South Parks Road, Oxford, OX1 3QZ, United Kingdom}
	
	\date{\today}
	
	\begin{abstract}
We introduce a scheme to design patchy particles so that a given target crystal is the global free-energy minimum at sufficiently low temperature. A key feature is a torsional component to the potential
that only allows binding when particles have the correct relative orientations. In all examples studied, the target crystal structures readily assembled on annealing from a low-density fluid phase, 
albeit with the simpler target structures assembling more rapidly. The most complex example was a clathrate with 46 particles in its primitive unit cell. We also explored whether the structural information encoded in the particle interactions could be further reduced. For example, removing the torsional restrictions led to the assembly of an alternative crystal structure for the BC8-forming design, but the more complex clathrate design was still able to assemble because of the greater remaining specificity.
	\end{abstract}
	\maketitle
	
	\section{Introduction}\label{sec:intro}

Our work aims to address a general, fundamental question: ``How can we design
particles such that they assemble into a given ordered structure?'' In
particular our focus is on the formation of a target complex crystal. Such a
question is, of course, of great importance to materials
design.\cite{Whitesides2002a,Talapin2010,Thorkelsson2015} As synthesis methods
to produce an increasing variety of complex particles, both in the nanoparticle
and colloidal domains, continue to
progress,\cite{Glotzer2007,Yi2013,Manoharan2015,Boles2016,Ravaine2017} it is
important to understand what properties such particles should possess in order
to form a specific structure, and hence have particular material properties.
For example, an oft-mentioned target is colloidal particles that can form a
diamond lattice, because of the potential favourable photonic properties.\cite{Maldovan2004}

Although the anisotropic shape of particles can be a powerful way of
controlling the structures into which particles assemble, and for which there
has been much recent progress,\cite{Damasceno2012,Geng2017} here we restrict
ourselves to particles that are spherical in shape. For such particles, there
are two main approaches to particle design that have been explored. Firstly,
there is the use of isotropic, and typically long-ranged, potentials with a
complex radial dependence (e.g.\ possessing features such as multiple minima
and repulsive shoulders). As the relationship between the features of these
potentials and the structures they adopt is often not obvious, one approach is
to map out the structures formed as a function of the parameter space of the
potential in search of interesting ordering
behaviour.\cite{Engel2007,Dotera2014,Engel2015} Another approach is to use
inverse design techniques to evolve a potential in order to stabilize a given
target
structure.\cite{Torquato2009b,Marcotte2013,Jain2013,Lindquist2017,Adorf2018}
Although both approaches have revealed a rich range of structural behaviour,
how to engineer particles to have such complex radial forms is not
straightforward. 

Secondly, there is the use of particles with directional attractive
interactions.  Typically, these are ``patchy'' particles where short-ranged
attractive interactions only occur when the patches on adjacent particles are
aligned.\cite{Bianchi2011}  One advantage is that the geometry of the patches
has a clear and direct relationship to the preferred local coordination shell
around a particle.  However, how these local environments join together to
determine the overall global structure may be less obvious. For example, there
are many potential structures that have local tetrahedral coordination.
Although exploring state points and different potential parameters (e.g.\ patch
width) have located conditions under which tetrahedral patch particles
generally form (cubic or hexagonal) diamond\cite{Romano2011} and even clathrate
structures,\cite{Noya19} competing structures are sometimes observed, and the
situation is somewhat unsatisfactory from a design point of view.  One approach
around this problem is to increase the number of particles and patch types in a
way that is only compatible with the target structure, as has recently been
been done for cubic and hexagonal diamond, albeit with assembly only being
succesful when seeded or grown from a template.\cite{Patra18} 

Another potential solution is to introduce a torsional component to the patchy
interactions, as this then allows the relative orientations of the two
particles to be fully determined, hence providing control of the
second-neighbour shell. For example, this approach has been exploited to create
tetrahedral patchy particles that readily form a cubic diamond
crystal.\cite{Zhang2005} Here, we further explore how the patchy particles with
torsionally-specific interactions can be used to form complex crystal
structures.  In particular, our aim is to come up with a general design scheme
that should work for any target crystal structure, however complex.  Previous
work on patchy particles and crystallization has mainly focussed on particles
without torsions. A significant fraction of that work has been on 2D
crystals,\cite{Doye2007,Doppelbauer2010,Antlanger2011,VanderLinden2012,Reinhardt2013,Whitelam2016,Chen18}
and those studies focussing on 3D crystals have mainly been with one-component
systems.\cite{Noya2007,Romano2010,Noya2010,Romano2011,Saika-Voivod2011,Dorsaz2012,Romano2012a}
By contrast, most work using torsionally-specific patchy particles has been on
the assembly of finite structures (analogous to protein
complexes),\cite{Wilber2009b,Villar2009} an exception being the study on
diamond mentioned above.\cite{Zhang2005} 
	
Of course, the potential down-side of introducing torsional interactions is
that this adds an extra level of complexity when trying to
experimentally realize such particles.  Perhaps the most directly analogous
experimental examples
have been spherical colloidal particles that have DNA origami belts wrapped
around them that have complementary patches.\cite{BenZion2017a} Although
globular proteins are of course not perfectly spherical, protein-protein
interactions mediated by complementary interacting patches of surface amino
acids are typically torsionally specific.  Such control allows proteins to
assemble into well-defined ordered structures. Most frequently these are
finite-sized protein complexes,\cite{Levy06d} perhaps the most iconic being icosahedral virus
capsids, but there are some examples of proteins that are designed to form 2D
and 3D crystals {\it in vivo}.\cite{Doye06b,Coulibaly07,Pum13,GarciaSeisdedos18} Furthermore,
although proteins have traditionally been hard to programme because of the
difficulty of predicting structure from sequence, great strides to overcome
this hurdle have been recently made, including the {\it de novo} design of
protein complexes and arrays.\cite{Huang16,Yeates17,Chen19}  

DNA nanotechnology perhaps provides the most promising avenue to realize
designable analogues of our model particles. Multi-arm DNA tiles have been
shown to form a wide range of polyhedra\cite{He2008,Zhang2008a,Zhang12c} and 2D
arrays\cite{Yan2003,He2005,He2006,Zhang2008a,Zhang2013,Zhang2016,Liu2019} that
bear striking similarities (e.g.\ equivalent networks of interactions) to those
formed by patchy particles.\cite{Doye2007,Wilber2009b,Reinhardt2016} Each tile
arm is made of two helices with the arms interacting via the hybridization of
single-stranded overhangs. Although the nature of the excluded volume is very
different from our model patchy particles, the assembly behaviour at
low-density in both cases is dominated by the directional attractions. A more
significant difference is the flexibility of the tiles with rigidification
occurring as a result of assembly into the target structure.\cite{Schreck16} At
a larger scale, three-arm DNA origami (where the angles between the arms are
constrained by bracing cross-struts) have been used to form a variety of
polyhedra.\cite{Iinuma14} Torsionally-specific assembly of DNA origami into
larger ordered structures can also be achieved through stacking interactions
between shape complementary surface features,\cite{Gerling15} and has been
utilized to form crystal lattices\cite{Zhang18} and large gigadalton
polyhedra.\cite{Wagenbauer17b}

	\section{Methods}\label{sec:methods}
	
	\subsection{Interaction potential}\label{sec:potential}
	
	The patchy-particle model we use derives from, but generalises, that used in Ref.~\onlinecite{Wilber2009b}. 
	The model particles interact with an isotropic repulsion, but their attraction depends on the relative orientation of the particles. The single-site, pairwise interaction potential between particles $i$ and $j$, $V_{ij}$, is given by
	\begin{widetext}
		\begin{equation}
		V_{ij}(\mathbf{r}_{ij},\mathbf{\Omega}_i,\mathbf{\Omega}_j) = \begin{cases}
		V_{\mathrm{LJ}}^\prime(r_{ij}) & : r_{ij} < \sigma_{\mathrm{LJ}}^{\prime} \\
		V_{\mathrm{LJ}}^\prime(r_{ij}) \underset{{\mathrm{patch~pairs~}\alpha,\beta}}{\max} \left[ \varepsilon_{\alpha\beta}V_{\mathrm{ang}}(\mathbf{\hat{r}}_{ij},\mathbf{\Omega}_i,\mathbf{\Omega}_j)V_{\mathrm{tor}}(\mathbf{\hat{r}}_{ij},\mathbf{\Omega}_i,\mathbf{\Omega}_j) \right] & : r_{ij} \geqslant \sigma_{\mathrm{LJ}}^{\prime}
		\end{cases},
		\label{eq:V}
		\end{equation}
	\end{widetext}
	where \rbij{} is the interparticle vector, 
$\alpha$ and $\beta$ are patches on particles $i$ and $j$ respectively, and $\mathbf{\Omega}_i$ is the orientation of particle $i$. 
	
	Equation~(\ref{eq:V}) is based on a cut-and-shifted Lennard-Jones (LJ) potential,
	\begin{equation}
	V_{\mathrm{LJ}}^\prime(r) = \begin{cases} V_{\mathrm{LJ}}(r)-V_{\mathrm{LJ}}(r_{\mathrm{cut}}) & : r < r_{\mathrm{cut}} \\
	0 & : r \geqslant r_{\mathrm{cut}}
	\end{cases},
	\end{equation}
	where
	\begin{equation}
		V_{\mathrm{LJ}}(r)=4\epsilon_{\mathrm{LJ}} \left[ \left( \frac{\sigma_{\mathrm{LJ}}}{r} \right)^{12} - \left( \frac{\sigma_{\mathrm{LJ}}}{r} \right)^{6} \right]
	\end{equation}
	is the standard LJ potential, \slj{} and \elj{} are, respectively, the interparticle distance at which $V_{\mathrm{LJ}}=0$ and the minimum value of $V_{\mathrm{LJ}}$.
	$\sigma_{\mathrm{LJ}}^{\prime}$ in Eq.~\ref{eq:V} corresponds to the distance at which $V_{\mathrm{LJ}}^\prime$ passes through zero. We set the cutoff distance \rcut{}~$=2.5$\,\slj{}.
	
	The overall potential (Eq.~\ref{eq:V}) has two regimes (but is continuous, by design). In the shorter-distance regime (\rij{}~$< \sigma_{\mathrm{LJ}}^{\prime}$), the potential is simply the isotropic LJ repulsion. In the longer-distance regime (\rij{}~$\geqslant \sigma_{\mathrm{LJ}}^{\prime}$) the LJ interaction is modulated by a dimensionless prefactor, \vepab{}, which is specific to the two patch types involved, and two orientationally dependent functions, \Vang{} and \Vtor{}. For any pair of particles $i$ and $j$, only the pair of patches $\alpha$ and $\beta$ that maximises \vepab{}\Vang{}\Vtor{} is considered to interact. In this work, \vepab{} only takes on binary values, i.e.~1 for patches that interact or 0 for patches that do not, but it could in principle be used to vary the strength of different patch-patch interactions.
In practice (see section~\ref{sec:design_approach}), we divide patches into \emph{types}, each of which has the same properties, and define a matrix of values of \vepab{} for all pairs of patch types.


	The \emph{angular} modulation term \Vang{} is a measure of how directly the patches $\alpha$ and $\beta$ point at each other, and is given by
	\begin{equation}
	V_\mathrm{ang}(\mathbf{\hat{r}}_{ij},\mathbf{\Omega}_i,\mathbf{\Omega}_j) = \exp \left( -\frac{\theta_{\alpha ij}^{2}}{2\sigma_{\mathrm{ang}}^{2}} \right) \exp \left( -\frac{\theta_{\beta ji}^{2}}{2\sigma_{\mathrm{ang}}^{2}} \right) .
	\label{eq:Vang}
	\end{equation}
	$\theta_{\alpha ij}$ is the angle between the patch vector $\mathbf{\hat{p}}_{i}^{\alpha}$, representing the patch $\alpha$, and $\mathbf{\hat{r}}_{ij}$. \sang{} is a measure of the angular width of the patch. 
	
	The \emph{torsional} modulation term \Vtor{} describes the variation in the potential as either of the particles is rotated about the interparticle vector \rbij{}, and is given by
	\begin{equation}
	V_{\mathrm{tor}}(\mathbf{\hat{r}}_{ij},\mathbf{\Omega}_i,\mathbf{\Omega}_j)
	= \exp\left(
		- \frac{1}{2\sigma_{\mathrm{tor}}^{2}}\left[
				\min\limits_{\phi^{\mathrm{offset}}_{\alpha\beta}}
				\left(\phi_{\alpha\beta}-\phi^{\mathrm{offset}}_{\alpha\beta} \right)
			\right]^{2}
	\right).
	\label{eq:Vtor}
	\end{equation}
	To define \phiab{}, a unique \emph{reference vector} (usually one of the other patch vectors) is associated with each patch. \phiab{} is then the angle between the projections of the reference vectors for patches $\alpha$ and $\beta$ onto a plane perpendicular to $\mathbf{\hat{r}}_{ij}$.
	Generalising the original model,\cite{Wilber2009b} we also define an \emph{offset angle} for each patch pair, \phiaboff{}. This allows us to specify a non-zero preferred 
	value of the torsional angle, \phiab{}, favouring some relative torsional rotation. More than one equivalent offset angle can be defined (in order to capture the site symmetry of the particle in the target structure), in which case we find the minimum value of \phiab$-$\phiaboff{} across the set of equivalent offset angles. \Vtor{} is maximised when the torsional angle $\phi_{\alpha \beta}$ matches one of the offset angles, 
	and twisting around the interparticle vector, away from \phiaboff{}, is penalised. Thus, the torsional component of the potential ensures bonded patches have the correct relative orientation. \stor{} is the torsional `patch width' and controls how quickly the potential energy increases on deviating from the preferred torsional alignment.
	
	As in Ref.~\onlinecite{Wilber2009b}, we fix the ratio of the two patch widths; we use \stor{}$= 2\,$\sang{}$= 0.6\,\mathrm{rad}$ throughout. 
	The results are relatively insensitive to the precise values (or ratio) of these parameters, 
	as long as the patches are sufficiently specific to favour the target structure and not too narrow that kinetic accessibility is hindered.\cite{Wilber2009b}
	Note, as \sang{}~$\to\infty$ and \stor{}~$\to\infty$, the isotropic LJ potential is recovered. We use \slj{} and \elj{} as our (reduced) units of length and energy, respectively, and use the reduced temperature $T^{*}=k_{\mathrm{B}}T/\epsilon_{\mathrm{LJ}}$.
		
	\subsection{Simulation method}\label{sec:simulations}
	
	To simulate our model, we perform standard Metropolis\cite{Metropolis1953} Monte Carlo (MC) simulations in the canonical ($NVT$) ensemble with (equally likely) single-particle translation and rotation moves. This leads to diffusive motion of the particles as is appropriate for colloidal particles in solution.
	We initiate each simulation in a random configuration with particle number density $\rho=0.1\,\sigma_{\mathrm{LJ}}^{-3}$ at a temperature corresponding to the stable low-density fluid phase. We use $N\approx1000$, normally choosing a cubic multiple of the number of particles in the target structure's unit cell. We anneal from the fluid phase to the two-phase region, corresponding to a coexistence of crystalline and very low-density fluid phases, lowering the reduced temperature at a rate of $2\times10^{-4}$ per $2.2\times10^{5}$~MC cycles. (Annealing is not continuous, but stepwise.) We repeat this protocol a number of times for each target structure.
	
	
	We chose to assemble large crystalline clusters from a low-density
	fluid, as this is typical of the conditions under which we would expect
	such patchy colloids 
	to be assembled. Note that we do not expect our particles to have a
	stable liquid phase, as the torsional component of the potential
	inhibits the formation of low-energy disordered configurations.\cite{Wilber2009b}
	
	\section{Design of patchy particles}\label{sec:design}
	
	\subsection{General approach}\label{sec:design_approach}
	
	To design a set of patchy particles that forms a target structure, we use the following scheme.
%
%
	For each particle in our target structure, we define patch vectors pointing at its nearest-neighbour coordination shell. We divide the patchy particles in the unit cell into \emph{types}, based on the structure's crystallographic symmetry. Particles of the same type have all the same properties (number of patches, patch vectors, and all patch properties). We then include the appropriate number of particles of each type in our simulation box.
	
	In general, particles are of the same type if they map onto each other by the symmetry operations of the structure's space group; but there is an exception. By way of explanation, the particles in the target structure's unit cell can be categorised by the \emph{Wyckoff positions}\cite{Wondratschek2006, Hammond2015} they occupy. Distinct particles occupying the same Wyckoff position map onto each other by the group's symmetry operations. Each occupied Wyckoff position corresponds to either one or two \emph{types} of patchy particle, in our scheme. For space groups without mirror planes or improper rotations, each occupied Wyckoff position corresponds to \emph{one} patchy-particle type. For space groups with mirror planes or improper rotations, occupied Wyckoff positions which lie on a mirror plane or improper rotation axis correspond to \emph{one} patchy-particle type; but other occupied Wyckoff positions correspond to \emph{two} patchy-particle types, since particles that map onto each other by mirror planes or improper rotations have enantiomeric environments. In the latter case, there are equal numbers of particles of each type, and the two patchy-particle types are enantiomeric---they are the same except for equal and opposite offset angles (\phiaboff{}) on all pairs of corresponding patches.
	
	

	As for particles, we categorise patches into \emph{types}. Patches of the same type have the same properties. 
	On a given particle, patches that are related by the site symmetry (at that Wyckoff position) are usually equivalent and so are of the same patch type. The exception is patches that are \emph{only} related by a mirror plane or improper rotation; these patches are of different patch types, but differ only in their offset angles, which are equal and opposite. Otherwise, patches are of distinct types.
	
	For each patch $\alpha$, we define \vepab{}~$=1$ for all patches $\beta$ to which $\alpha$ points in the target structure. For all other $\beta$, we define \vepab{}~$=0$; these patch pairs do not interact in the target structure. For each patch, we choose a reference vector from one of the other patch vectors on that particle; where possible the reference vector chosen lies on a symmetry axis or plane. For each complementary (i.e.~interacting) patch pair $\alpha\beta$, we calculate the offset angle(s) \phiaboff{} in the target structure. An offset angle is given by the torsional (or dihedral) angle for a sequence of four coordinating particles: the two particles interacting via their patches, and the particles to which the two corresponding reference vectors point. If the reference vector is a patch vector for which other patches on that particle are of the same patch type, then any of these symmetry-equivalent patches could have been used as the reference vector; to capture this symmetry when using a single reference vector we allow multiple offset angles, one for each of the equivalent possible reference vectors. Thus each offset angle corresponds to a symmetry-equivalent orientation of the particle in the target structure.
	

	The patchy interactions specified by the above rules \emph{fully}
	specify the target structure, and thus are sufficient to make it the
	potential energy global minimum for a system with the correct
	composition of the different particles. The torsional interactions are
	key to this, as they ensure that the first coordination shell of a
	particle not only bonds in the correct relative positions but that
	these particles also have the correct relative orientations, thus
	ensuring that the next coordination shell also binds in the correct
	positions. For example, without torsions, it would be hard to design a
	patchy-particle system that selectively forms cubic rather than
	hexagonal diamond (or {\it vice versa}) since the structures only
	differ in their second coordination shell. 

	In this scheme, we use symmetry to try to minimize the amount of information that needs to
	be encoded in the particles. It is an interesting question whether one can reduce this information further 
	whilst still retaining robust structural selectivity---we start to explore this in Section \ref{sec:results_information}.
	Also, in this paper we have restricted ourselves to target structures for which the inter-particle distances between bonded particles are very similar (the difference between the longest and shortest bonds in the ClaI, ClaII, BC8, and A15 structures that we consider in Section \ref{sec:results} are $2.1$\%, $1.4$\%, $3.7$\%, and $11.8$\%, respectively, and all bonds in cP4 are of the same length), 
	allowing us to use particles of all the same size. However, there is no reason why the approach could not be extended to structures where different-sized particles would be required, as long
	as the bond distances are sufficiently additive.

	\subsection{Application to example structure}\label{sec:design_examples}
		
\begin{figure*}
		\centering
		\includegraphics[scale=0.95]{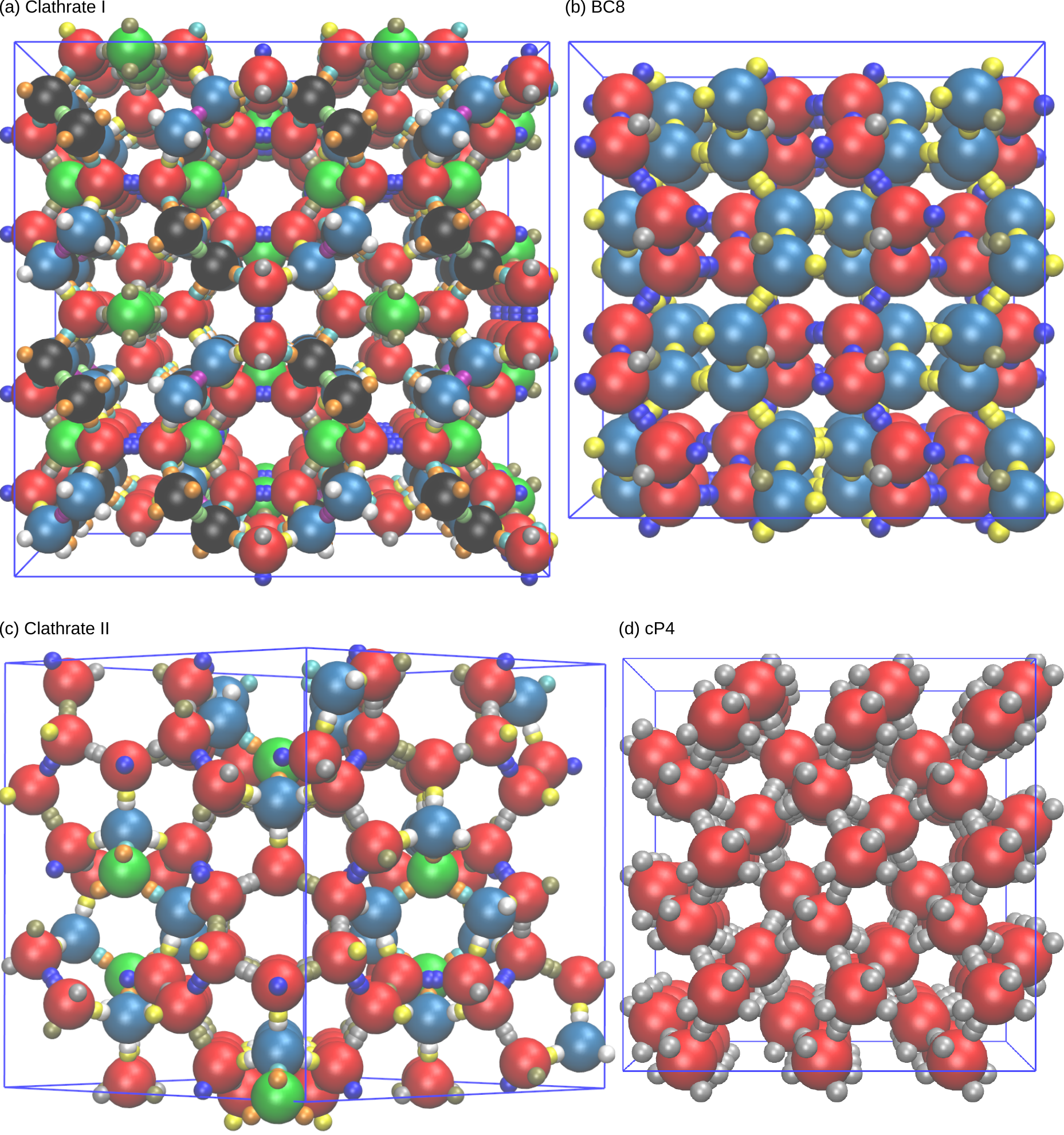}
		\caption{\label{fig:crystals}Crystal structures of the (a) clathrate type I, (b) BC8, (c) clathrate type II, and (d) cP4 targets, depicted with patchy particles. 
		The patch vectors are represented by the small particles on the surface of the main particles. The particle and patch sizes have been chosen for clarity.
		Different colours correspond to different types (for both particles and patches). Details of each structure are given in Table~S1; particle designs are specified in Table~S2 and pictured in Fig.~S1.}
	\end{figure*}

	Here we illustrate the above scheme 
	for one of our target structures,
	namely clathrate type I (ClaI), which is shown in Fig.~\ref{fig:crystals}(a). The supplementary material contains further information on this structure (Table~S1) and the set of patchy particles designed to form it (Table~S2 and Fig.~S1(a--d)).
	
	ClaI is in space group Pm$\overline{3}$n (number 223) and has 46 particles in its unit cell, located at three Wyckoff positions: 24k (red in Fig.~\ref{fig:crystals}(a)), 16i (blue and black), and 6d (green). (The number associated with a Wyckoff position denotes how many particles occupy this position in the unit cell.) 
	Positions 24k and 6d lie on mirror planes, so each corresponds to a single patchy-particle type, whereas position 16i does not, so corresponds to two enantiomeric particle types. Thus we define four types of patchy particles, with 8 particles in the unit cell for each of the two 16i types.

	
	All particles have a coordination number of four and so have four patches. On the 24k particles, one patch interacts with a patch on another 24k particle, one with a 6d particle, and one with each of the two types of 16i particles. The two former patches lie on a mirror plane and are unique. The latter two patches map onto each other only by a mirror plane, and so are of different types, with equal and opposite offset angles. 
	For all these patches, we choose as reference vectors one of the two patch vectors that lies on the mirror plane passing through this site. The four patches on the 6d particles each interact with a patch on a (different) 24k particle, and map onto each other by the site symmetry ($D_{2d}$). These patches are all of the same type, and the reference vector for each patch is chosen to be the vector for the patch that is related by both a 2-fold rotation and a mirror plane. For the 16i particles, one patch is unique and interacts with an equivalent 16i particle.
	The other three patches interact with 24k particles, and map onto each other by a 3-fold rotation, so are therefore of the same type. The patch vectors of any of the three symmetry-related patches could be used as the reference vector for the unique patch; we choose one but have three equivalent offset angles to account for the symmetry. We use the unique patch, which lies on a 3-fold axis, as the reference vector for the other three patches. The two types of 16i particle are identical except for having equal and opposite offset angles for all patches.
	
	\section{Self-assembly simulation results}\label{sec:results}
	
	We applied our patchy-particle design scheme to the five target structures shown in Figs.~\ref{fig:crystals}(a--d) and \ref{fig:A15}(a): clathrate type I (ClaI), BC8, clathrate type II (ClaII), cP4 and A15. The structures are relatively complex (i.e.~have a large number of particles in their unit cells), mostly open (i.e.~have a low packing fraction) and have features, e.g.\ the pores in the clathrates, that may be of interest for functional materials. 
	ClaI, BC8, and ClaII are three examples from the many structures that involve just tetrahedral coordination. As mentioned in the Introduction, previous work on particles with tetrahedral patches \cite{Zhang2005, Romano2010, Noya2010, Romano2011, Saika-Voivod2011, Dorsaz2012,Noya19} has shown that it is hard to control the assembly product without torsional interactions.
	Also, ClaI, BC8, ClaII, and cP4 have been found to be stable in potential parameter space near to an icosahedral quasicrystal (IQC) for a one-component system with an isotropic potential with multiple minima.\cite{Engel2015} 
	A15 provides an example crystal with a higher average coordination number.
	
	The supplementary material contains further information on these structures (Table~S1), and the sets of patchy particles designed to form each (Table~S2 and Fig.~S1).

	\subsection{General assembly kinetics}\label{sec:results_general}
	
\begin{figure}
		\includegraphics[scale=0.66,trim=0 12 0 0, clip]{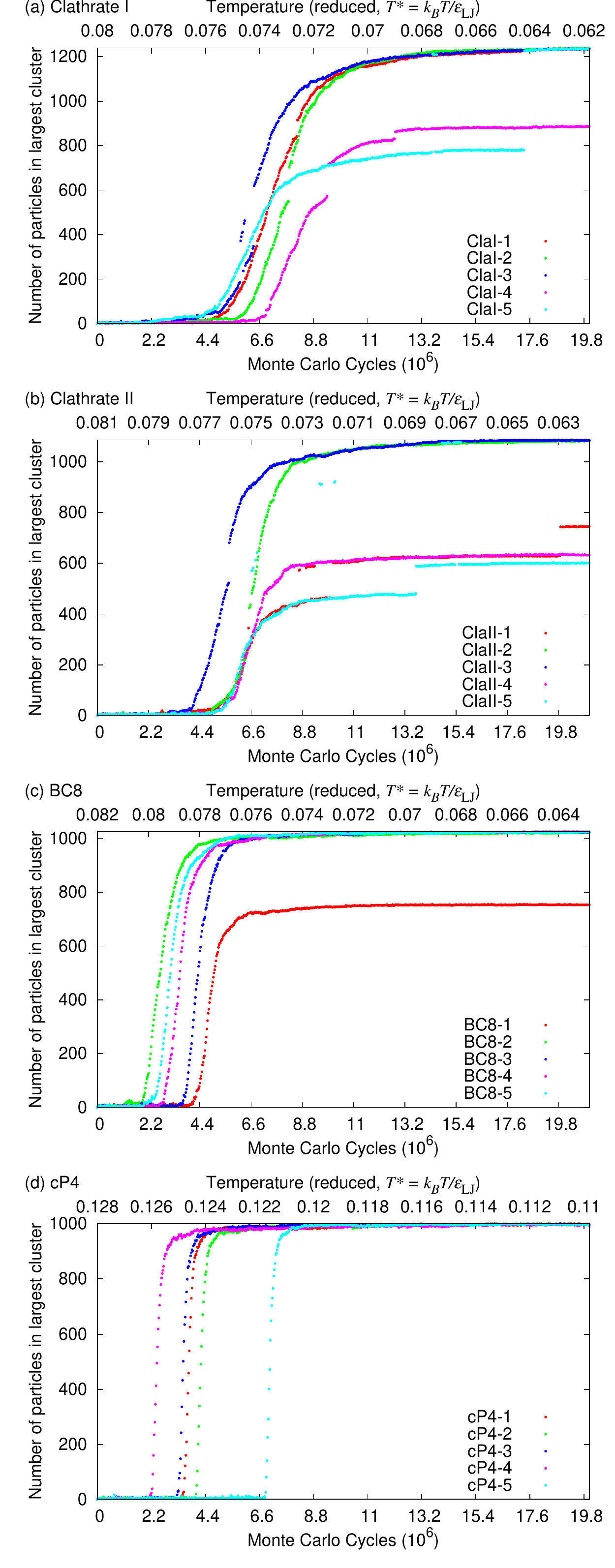}
		\caption{\label{fig:largest_clusters}Cluster nucleation and growth during annealing simulations for the systems designed to form (a) clathrate I, (b) clathrate II, (c) BC8, and (d) cP4 target structures. Each plot shows the crystal growth, measured in terms of the largest cluster, against progress in the simulation, measured both in in terms of the number of MC cycles (bottom axis) and the reduced temperature (top axis). For each design, the results
of five distinct simulations are presented.
The maximum value on the vertical axis in each plot corresponds to the total 
number of particles in the simulation. 
The annealing rate was the same for all simulations.
}
	\end{figure}

\begin{figure}
		\includegraphics[scale=0.97]{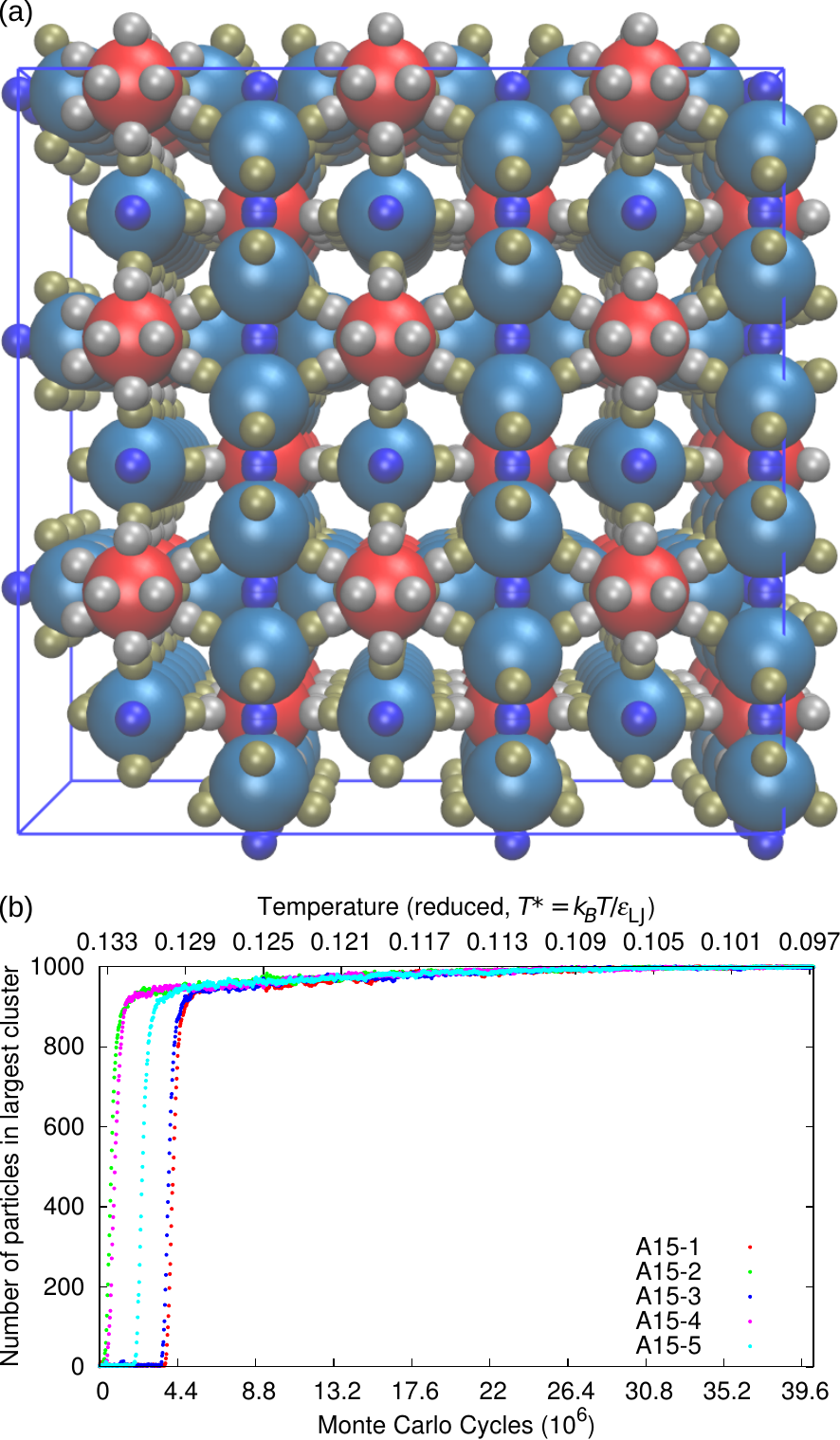}
		\caption{\label{fig:A15}(a) Crystal structure (as for Fig.\ref{fig:crystals}) and (b) cluster nucleation and growth (as for Fig.~\ref{fig:largest_clusters}) for the A15 target.}
	\end{figure}

	All five target structures successfully assembled from their respective patchy-particle systems. To characterize the nucleation and growth of a crystalline cluster from the dilute fluid phase, we use a simple order parameter: the number of particles in the largest cluster in the system, where we define a \emph{cluster} as any network of sequentially bonded particles, where two particles are \emph{bonded} if 
	their pairwise interaction energy is less than $-0.2$\,\elj{}. 
	In Figs.~\ref{fig:largest_clusters}
	and \ref{fig:A15}(b) this quantity is plotted over the course of the simulation for each of the five repetitions for each structure. They show that by the end of most simulations, the largest cluster contained all or almost all of the particles in the simulation box. When this did not occur, it was simply because one (or sometimes two) additional, smaller crystalline cluster(s) had formed. Visual inspection of the configurations confirms the target structure correctly formed in all cases---both for the largest clusters, and any smaller clusters. 
	
\begin{figure*}
		\centering
		\includegraphics[scale=0.93]{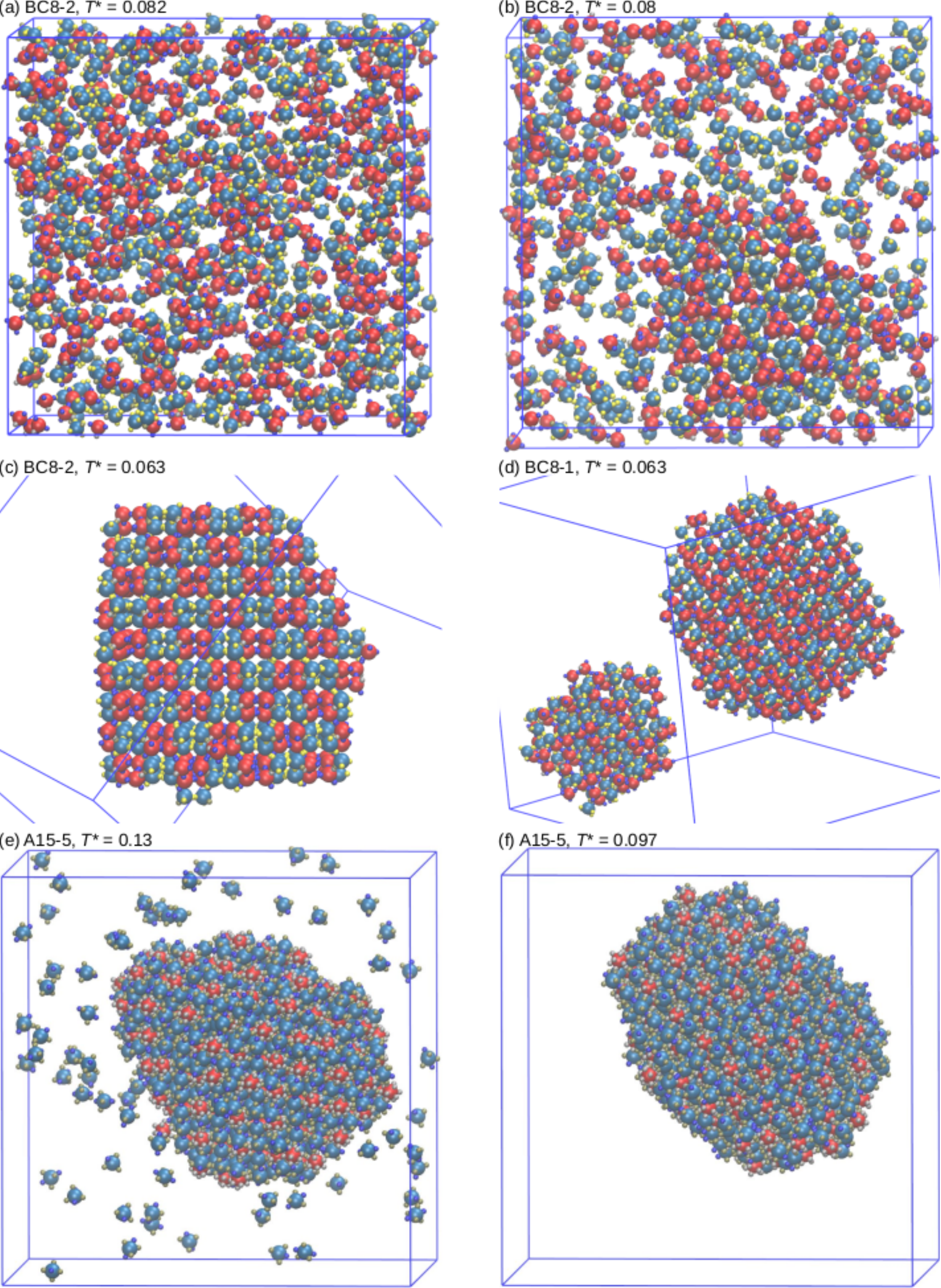}
		\caption{\label{fig:stages}Simulation snapshots showing: (a)--(c) three stages in the self-assembly of BC8 during repetition 2 ((a) fluid phase, (b) nucleation of a crystalline cluster (towards the bottom right) within the fluid phase, and (c) final cluster), (d) the final state of BC8 in repetition 1 (Fig.~\ref{fig:largest_clusters}(c)), and (e,f) the final stages of the assembly of the A15 target during repetition 5 (Fig.~\ref{fig:A15}(b)).}
	\end{figure*}

	The assembly behaviour was generally as hoped for in the design process. When the temperature was sufficiently below the 
fluid-crystal binodal,
assembly starts to occur. Snapshots from the self-assembly of BC8, as an example, can be seen in Fig.~\ref{fig:stages}(a--c): (a) shows the low-density fluid phase, (b) the nucleation of a cluster, and (c) the final cluster, containing all particles. Nucleation appears consistent with classical nucleation theory.
	Assembly progresses by the addition of monomers (and occasionally small clusters) onto growing clusters. Disordered aggregates, incorrect structures, and other kinetic traps were not observed. Their absence is a feature of assembly when torsional interactions are included.\cite{Wilber2009b}
	
	There is a narrow temperature window for good nucleation and crystal
	growth from the fluid, in which a single, large cluster forms, rather
	than multiple smaller clusters. Within the two-phase region 
	at \emph{high} temperatures, the free energy barrier to
	nucleation is high, so it does not occur on practical timescales.
	By contrast, at \emph{low} 
	temperatures, the free energy barrier is low, and so multiple nucleation
	occurs and many clusters grow. 
	These clusters are unlikely to have complementary shapes so as to
	fit together to form a defect-free larger cluster. 
	However, neither can these clusters easily break apart and
	reassemble, because the low temperature offers limited thermal energy
	to break bonds. Thus, the system is kinetically trapped, and tends to retain
	separate clusters, rather than forming a single large cluster. For example, this can be seen
	in Fig.~\ref{fig:stages}(d) for the BC8 design. 
	Instead, assembly best occurs in an \emph{intermediate}
	temperature window, where nucleation can occur but does so only rarely, and the rate of growth is much greater than the
	rate of nucleation. Consequently once a cluster does nucleate, it grows rapidly,
	and it is unlikely other clusters will nucleate in this period. 
	
	In our preliminary simulations, we used a faster annealing rate. Consistent with the above discussion, nucleation 
	typically occurred at lower temperatures, and the formation of multiple clusters was more likely.
	Nonetheless, these quicker simulations helped guide the choice of temperature range for the production runs.
		
	In Figs.~\ref{fig:largest_clusters} and \ref{fig:A15}(b), among repeated simulations for the same design, clusters nucleated at a range of temperatures; this reflects the  
	stochastic nature of the nucleation process. 
	Multiple cluster formation is clearly more likely for those simulations, in which nucleation happens to occur at a lower temperature.
	For example, the BC8 repetition for which nucleation occurred at the lowest temperatures was the only one in which the largest cluster did not contain all particles at the end of the simulation
	(the final configuration is shown in Fig.~\ref{fig:stages}(d)).
	Multiple cluster formation is also more likely for those systems that have a slower growth rate (e.g.\ ClaI and ClaII).  
	In some cases involving multiple nucleation, the largest cluster combined with other clusters, as manifested in discontinuous jumps in its size (e.g.~ClaI-4 at $T^{*}\approx0.0715$ and $T^{*}\approx0.069$ (Fig.~\ref{fig:largest_clusters}(a))). When two separate clusters join, the number of initial intercluster bonds may be relatively few, and 
	so the joined clusters may soon break apart (e.g.~ClaII-5 twice briefly forms a larger cluster of $\sim900$~particles at $T^{*}\approx0.072$ (Fig.~\ref{fig:largest_clusters}(b))).
	
	\subsection{Differences between target structures}\label{sec:results_differences}
	
	It is apparent in Figs.~\ref{fig:largest_clusters} and \ref{fig:A15}(b) that some structures form more rapidly and successfully than others. cP4 forms most easily: in all simulations, the final largest structure contained all particles, and the cluster grew rapidly.
	Similarly, A15 forms easily, and BC8 forms almost as easily, with multiple clusters forming in only one instance (repetition 1). The reason these structures form more easily is most likely that their patchy-particle designs are simpler: there are fewer particles in the unit cell (4 for cP4, 8 for A15, 8 for BC8), fewer particle types (1, 2 and 2, respectively),
	and fewer patch types (1, 3 and 4). Thus, for these structures, there are fewer distinct configurations for the system to explore, and it can find the correct ones more quickly. In contrast, for both clathrates, at the end of many simulations the largest cluster did not contain all particles, and cluster growth was slower. ClaII and ClaI have, respectively, 34 and 46 particles in their unit cells, and their patchy-particle designs have 3 and 4 types of particles, and 7 and 9 types of patches. A system has more configurations to explore to find these more complex structures.	
		
	Whereas particles in ClaI, ClaII, and BC8 have 4 patches per particle, those in cP4 and A15 have more (cP4 has 6 patches per particle, and A15 particles has on average 7.5 patches per particle).
	Structures containing particles with more patches will be energetically more stable, and have higher melting points. This can be seen by the different nucleation temperatures in Figs.~\ref{fig:largest_clusters} and \ref{fig:A15}(b): the maximum nucleation temperatures (across all five repetitions, and defined as the first point at which the largest cluster always exceeds 10 particles) were $T^{*}\approx0.078$ for the ClaI and ClaII designs, $T^{*}\approx0.081$ for BC8, $T^{*}\approx0.126$ for cP4, and $T^{*}\approx0.133$ for A15.
	These temperatures are roughly in proportion to the (average) number of patches per particle in each structure.
	The higher nucleation temperatures for cP4 and A15 are also likely to aid assembly, as individual bonds are easier to break, thus facilitating particle rearrangements 
	to find the correct structure.

		
	

	That the two types of patchy particles in our design for A15 have different numbers of patches (one has 12 patches and the other 6) 
	gives rise to assembly behaviour that is somewhat different from the other designs. Assembly occurs in two `stages', as shown in Fig.~\ref{fig:stages}(e) and (f). First, a cluster forms that incorporates all the 12-patch particles, but only some of the 6-patch particles---all the interior sites for the 6-patch particles are occupied, but not all the surface sites---and so the assembled cluster is surrounded by 
	a gas of unbonded 6-patch particles (Fig.~\ref{fig:stages}(e)). Then, as the temperature is further decreased, the unbonded 6-patch particles gradually join onto the outside of the cluster (Fig.~\ref{fig:stages}(f)). This behaviour can also be seen in Fig.~\ref{fig:A15}(b): after the steep period of cluster growth, the curves do not become flat (as they do for most other structures), but have a slightly positive gradient. In this period, the largest cluster's size continues growing as more 6-patch particles add on. 
	The reason for this behaviour is simply the considerably smaller binding energy of the 6-patch particles.
	
	
		
	\subsection{Reduced structural information}\label{sec:results_information}

	The patchy-particle design protocol that we have introduced encodes
	\emph{sufficient} information in the particles to ensure the target
	structure both is the low-temperature free-energy global minimum, and,
	at least for all the examples considered above, is able to correctly
	assemble on annealing. Furthermore, we have exploited the symmetry of
	the target to try to minimize the number of particle and patch types.
	However, can one further reduce the encoded information whilst retaining the favourable assembly behaviour?
	For example, could one use fewer particle types, patches or patch types, and less (or no) torsional restrictions? 
	The potential advantage of such information reduction is that simpler designs are likely to be easier to realize experimentally.

	The danger of reducing the encoded information is, of course, that it is likely to increase the energetic stabilisation of alternative structures. 
	These competing structures may lead to a reduction in the kinetic accessibility of the target (due to them acting as kinetic traps) or may even become thermodynamically more stable than the target.
	For example, there are many structures with near-tetrahedral local coordination, and the less information is encoded in the design to specify one over the others, the more competition there will be between them. Thus, whether, and to what extent, the information in the particle design can be reduced is likely to be target dependent. 
	We begin to explore these questions through the examples below.
	
	\subsubsection{Fewer patches}\label{sec:results_information_patches}
	
	Intuitively, structures with high coordination environments (e.g.~A15) are good candidates for reducing structural information in their designs. Their large number of patchy bonds may over-specify the target structure, and, hence, there may be redundancy in their designs. Indeed, in our original A15 design we chose to ignore some neighbour information: it includes all 2a-6c (the labels refer to particle Wyckoff positions, as in Table~S2) nearest-neighbour bonds (4 per 6c particle, 12 per 2a particle; bond length $2.544$\,\r{A} in Cr$_{3}$Si) and the shorter 6c-6c nearest-neighbour bonds (two per 6c particle; $2.275$\,\r{A}), but not the slightly longer 6c-6c bonds (8 per particle; $2.786$\,\r{A}). Patches for the former two bond types
were sufficient to yield A15, 
but can the information content be reduced further still?
	
\begin{figure}
		\includegraphics[scale=0.98]{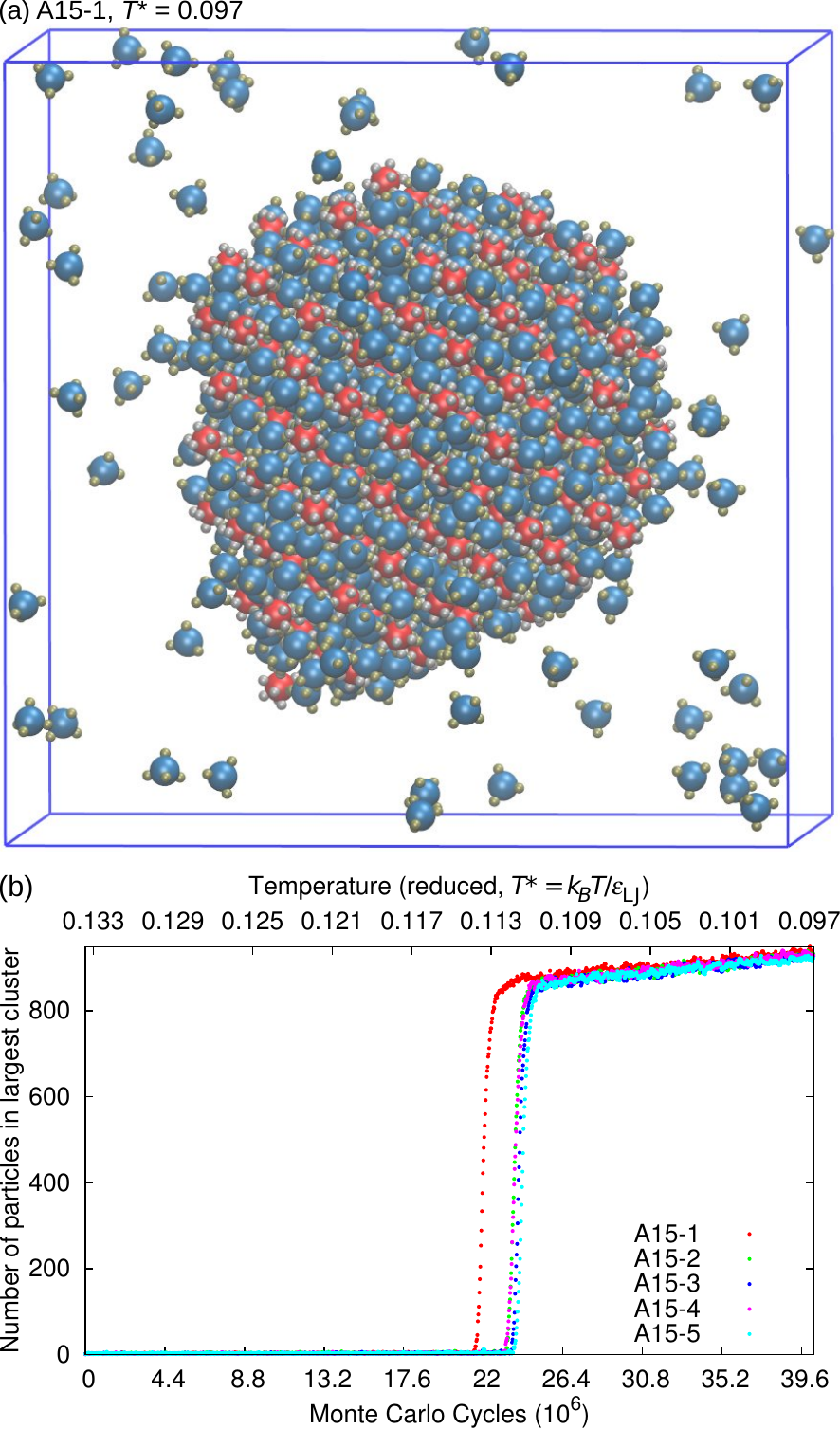}
		\caption{\label{fig:A15_lesspatches}Simulations of a system designed to form A15 but with a reduced number of patches: (a) example final configuration and (b) cluster-growth plot (as in Fig.~\ref{fig:largest_clusters}).}
	\end{figure}

	We designed a system with no 6c-6c patches, only patches for 2a-6c
	bonds: the design is the same as that in Table~S2 and Fig.~S1, except
	6c particles have only 4 patches, not 6 (patch numbers 5 and 6, both of
	type 3, are omitted). This system correctly formed A15 (all simulation
	details were the same as before). The results for this test are shown
	in Fig.~\ref{fig:A15_lesspatches}. As expected, crystallization
	occurred at a somewhat lower temperature than in the original
	simulations 
	($T^{*}\approx0.113$, previously
	$T^{*}\approx0.132$). Like with the original A15 design, assembly
	occurs in two stages, but it is now even more pronounced because of the
	bigger difference in the number of patches between the two particle
	types. Indeed at the end of the simulations ($T^{*}=0.097$), the
	surface 6c particles have still not all bonded to the cluster
	(Fig.~\ref{fig:A15_lesspatches})

The key feature that allows the reduced design to assemble correctly is that the network of remaining bonds is still able to fully define the target structure.
	
	\subsubsection{No torsional interactions}\label{sec:results_information_torsional}

	Another approach to reduce the information embedded in the particle design that we tested was to eliminate the torsional component of the interaction potential (i.e.~setting \Vtor{}~$=1$). This approach is attractive, given that torsional interactions are a potentially difficult aspect of our model to implement experimentally. However, such a change gives up direct control of the next-neighbour shell that is key to the generality of the success of the current patchy-particle design protocol, and thus has the potential to detrimentally affect the kinetics and thermodynamics of target assembly.
For instance, it is difficult for cubic diamond to form from tetrahedral patchy particles without torsional interactions;\cite{Zhang2005, Romano2010, Noya2010, Romano2011, Saika-Voivod2011, Dorsaz2012} however it is possible under some conditions (in particular, by using a seed\cite{Zhang2005, Noya2010} or via more selective interactions\cite{Dorsaz2012}).
	
	\begin{figure*}
		\centering
		\includegraphics[scale=0.98]{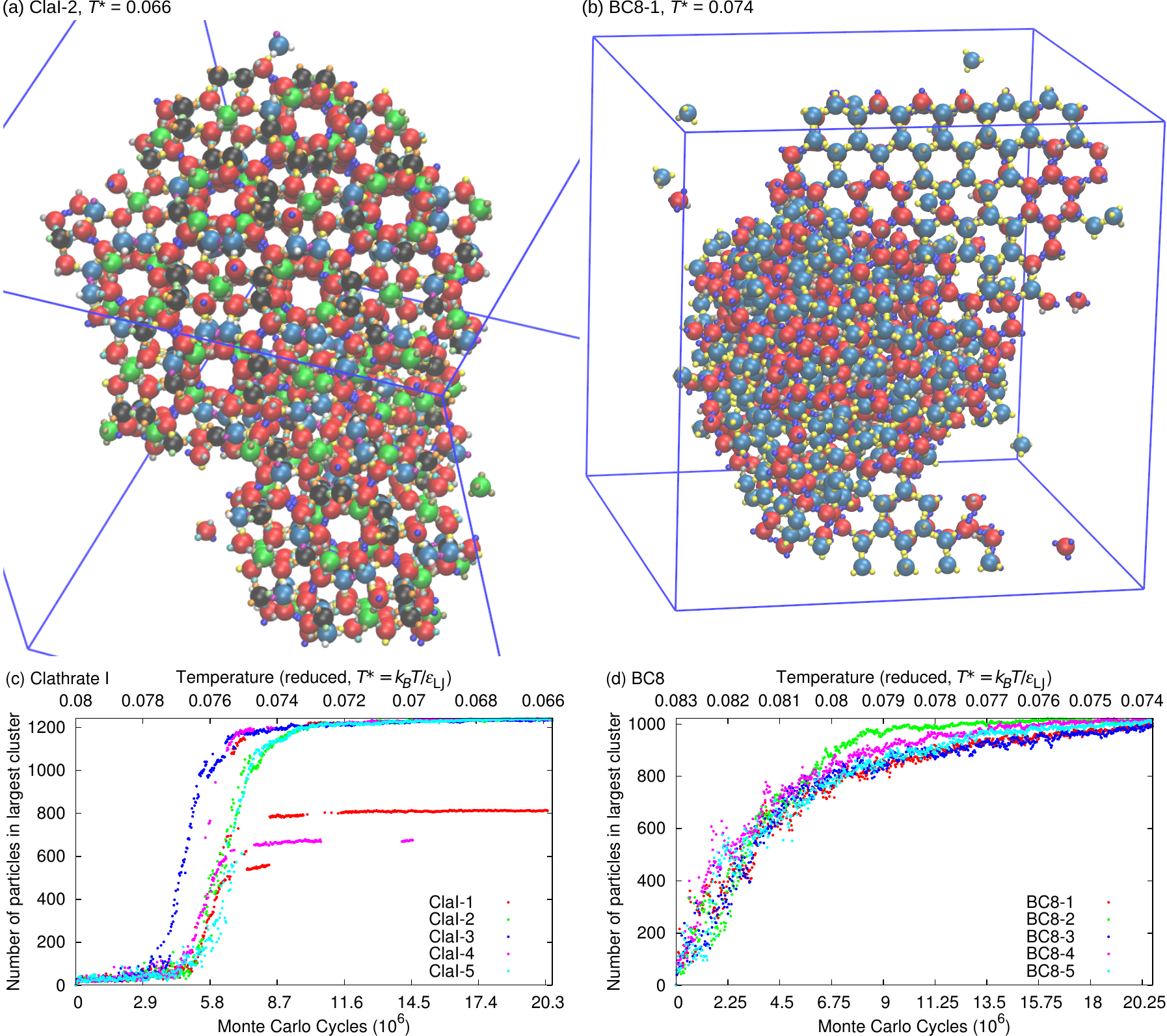}
		\caption{\label{fig:no_torsions}(a) and (b) Final configurations and (c) and (d) cluster-growth plots (as in Fig.~\ref{fig:largest_clusters}), in simulations of the sets of patchy particles designed to form (a,c) clathrate I and (b,d) BC8, but for which the torsional component of the potential was not included. Note the hexagonal diamond structure at the top and bottom of (b), connected by a disordered liquid-like droplet.}
	\end{figure*}
	
	To this end, we simulated designs with no torsional interactions (but with all other design details the same) for two sample structures, ClaI and BC8. The annealing rates were slightly faster ($T^{*}$ decreased by $2\times10^{-4}$ every $2.9\times10^{5}$~MC cycles for ClaI and $2\times10^{-4}$ every $4.5\times10^{5}$~MC cycles for BC8) and the temperature ranges narrower. The results for these simulations are shown in Fig.~\ref{fig:no_torsions}. ClaI assembled with roughly equal success compared to when torsional interactions were included: a sample final configuration is shown in Fig.~\ref{fig:no_torsions}(a); and the cluster-growth plot in Fig.~\ref{fig:no_torsions}(c) is similar to Fig.~\ref{fig:largest_clusters}(a). 
	However, BC8 did not. Instead, a disordered liquid-like cluster first nucleated from the low-density fluid. On further cooling, a different ordered structure, namely hexagonal diamond, nucleated from the liquid-like phase. An example final configuration is shown in Fig.~\ref{fig:no_torsions}(b). The hexagonal diamond crystallites were anisotropic in shape, due to its layered structure (the two particle types occupy alternating layers). In hexagonal diamond, the patches are not perfectly aligned (they were optimised for BC8), and so the energy of hexagonal diamond is somewhat higher than that of BC8. Therefore, the former is likely to be a kinetic product that is thermodynamically less stable. There was also a degree of variability in the results of the simulations: sometimes, liquid-like regions remained (as in Fig.~\ref{fig:no_torsions}(b)); sometimes, BC8 motifs were present. 
	
	Growth of the largest cluster for the BC8 design without torsional interactions (Fig.~\ref{fig:no_torsions}(d)) is slower and less smooth than with torsional interactions (Fig.~\ref{fig:largest_clusters}(c)). 
	This is simply because the initial cluster growth is of liquid droplets. Nucleation of multiple droplets initially occurs, followed by Ostwald ripening leading to the growth of the largest cluster by evaporation and re-condensation, and by cluster-cluster aggregation. The fluctuations in the cluster size reflect both the stochastic nature of the former, and that recently-joined clusters often break apart.

	These two examples illustrate that the necessity of torsional interactions is target dependent. In the more complex ClaI system, the specificity inherent in the multiple particle and patch types is sufficient to ensure that the target structure is both thermodynamically and kinetically favoured. However, removing the torsions from the simpler BC8 system allows a second fully-bonded hexagonal diamond crystal structure, which is more kinetically accessible than the BC8 target. Removing the torsions will also tend to stabilize the liquid phase,\cite{Wilber2009b} but whether this is detrimental to assembly is likely to be system specific. 
	
	\section{Conclusion}\label{sec:conclusion}
		
	We have demonstrated a systematic scheme to rationally design patchy particles that assemble into a given periodic structure. The scheme ensures the target structure is the free-energy global minimum structure below some cutoff temperature, with no nearby competing structures. 
	We have shown the scheme to be robust in enabling the formation of the target crystal for a variety of complex structures. 
	The specificity encoded by precise patch positions, patch interaction selectivity, and restrictions on torsional orientations programme the system to assemble into the correct structure. 
	Torsional interactions are a key feature of our designs, as they only permit the particles to bind together in a way that is consistent with the target structure, thus removing the possibility of  competing forms and ensuring the kinetic accessibility of the target. 

	Given that the experimental realization of these patchy particle
	designs would be challenging, it would be useful to know the minimal
	information that needs to be encoded in a patchy design to form a
	target structure. We have explored two potential ways to simplify
	the designs. In our general design scheme we introduce patches directed
	at all neighbouring particles in the target structure. However, for
	structures involving higher coordination numbers they may not all be
	necessary, as a design with a reduced number of interactions may still
	fully determine the target structure. This was the case for our
	simplified A15 design which still assembled successfully. The second
	simplification that we considered was the removal of the torsional
	specificity of the interactions. Perhaps surprisingly, the clathrate-I
	design was still able to assemble; the remaining specificity in this
	relatively complex design was still sufficient to favour correct
	assembly. However, the BC8 design no longer reliably assembled, because
	other competing structures were now compatible with the particle
	design; in such cases perhaps reintroducing greater specificity through
	increasing the number of patch or particle types may enable the target
	structure to form without torsions.\cite{Dorsaz2012,Patra18} These examples
	illustrate that there is unlikely to be a general scheme that enables
	the ``simplest'' design to be obtained.

	An interesting next challenge would be to obtain design rules for patchy particles to form (\emph{aperiodic}) target structures, such as quasicrystals. In two dimensions, patchy particles that can form dodecagonal\cite{VanderLinden2012, Reinhardt2013, Reinhardt2016} and metastable octagonal and decagonal\cite{Gemeinhardt2019} quasicrystals have been identified, but the formation of a three-dimensional aperiodic structure, such as an icosahedral quasicrystal, is a considerably 
	more difficult undertaking. 
	One of the issues is that quasicrystals can involve a mixture of order and disorder, and so there is not a single target structure; indeed, they can be stabilized by the resulting entropy.\cite{Reinhardt2013}
	The current study can potentially help as a starting point by providing the design rules necessary for crystalline approximants to quasicrystals; modifications to these designs that somewhat reduce their specificity may be a productive path to explore.

	\begin{acknowledgments}
		D.F.T. is grateful for funding via the ESPRC Centre for Doctoral Training in Theory and Modelling in Chemical Sciences, under grant EP/L015722/1.
		E.G.N. acknowledges funding from the Agencia Estatal de Investigaci\'{o}n (AEI) and the Fondo Europeo de Desarrollo Regional (FEDER) under grant numbers FIS2015-72946-EXP(AEI) and FIS2017-89361-C3-2-P(AEI/FEDER,UE), and the European Union’s Horizon 2020 research and innovation programme under the Marie Sk\l{}odowska-Curie grant agreement No.\ 734276.
		We thank Lorenzo Rovigatti for helpful advice about the simulation code.
	\end{acknowledgments}
	
	\section*{References}\label{sec:references}

%
	
\end{document}